\documentclass[runningheads]{llncs}
\usepackage{graphicx}
\usepackage{lipsum}
\graphicspath{{Figures/}}
\usepackage{caption}
\usepackage{subfig} 
\usepackage{amsmath}
\usepackage{amssymb}
\usepackage{epsfig}

\begin{document}
%
\title{An Advert Creation System for \\Next-Gen Publicity }
%

\author{
Atul Nautiyal\thanks{Authors contributed equally and arranged alphabetically.}\inst{1} \and
Killian McCabe$^{\star}$\inst{1} \and
Murhaf Hossari$^{\star}$\inst{1} \and 
Soumyabrata Dev$^{\star}$\inst{1} \and 
Matthew Nicholson\inst{1} \and  
Clare Conran\inst{1} \and 
Declan McKibben\inst{1} \and 
Jian Tang\inst{3} \and 
Xu Wei\inst{3} \and 
Fran\c{c}ois Piti\'e\inst{1,2}
}
%
%

\institute{The ADAPT SFI Research Centre, Trinity College Dublin \and
Department of Electronic \& Electrical Engineering, Trinity College Dublin \and
Huawei Ireland Research Center, Dublin\\
}

\authorrunning{A. Nautiyal, K. McCabe, M. Hossari, S. Dev et al.}
\titlerunning{An Advert Creation System for Next-Gen Publicity}

\maketitle              
\begin{abstract}
With the rapid proliferation of multimedia data in the internet, there has been a fast rise in the creation of videos for the viewers. This enables the viewers to skip the advertisement breaks in the videos, using ad blockers and `skip ad' buttons -- bringing online marketing and publicity to a stall. In this paper, we demonstrate a system that can effectively integrate a new advertisement into a video sequence. We use state-of-the-art techniques from deep learning and computational photogrammetry, for effective detection of existing adverts, and seamless integration of new adverts into video sequences. This is helpful for targeted advertisement, paving the path for next-gen publicity.

\end{abstract}
\section{Introduction}
\vspace{-0.1in}
With the ubiquity of multimedia videos, there has been a massive interest from the advertisement and marketing agencies to provide targeted advertisements for the customers. Such targeted advertisements are useful, both from the perspectives of marketing agents and end users. The advertisement agencies can use a powerful media for marketing and publicity; and the users can interact via a personalized consumer experience. In this paper, we attempt to solve this by designing an online advert creation system for next-gen publicity. We develop and implement an end-to-end system for automatically detecting and seamlessly changing an existing billboard in a video by inserting a new advert. This system will be helpful for online marketers and content developers, to develop video contents for targeted audience. 

\begin{figure}[htb]
\centering
\includegraphics[width=0.3\textwidth]{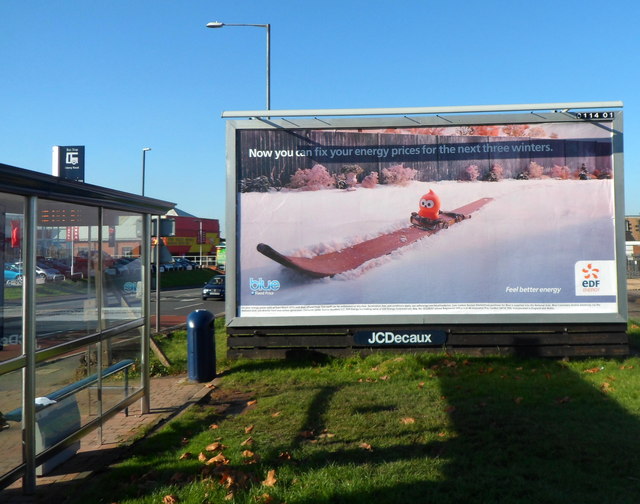}
\includegraphics[width=0.3\textwidth]{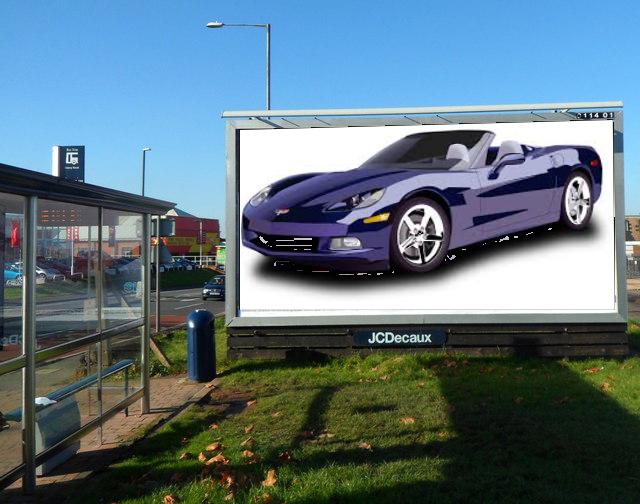}
\caption{New advert integrated into the scene at the place of an existing billboard.}
\label{fig:advert-story}
\vspace{-0.8cm}
\end{figure}

Figure~\ref{fig:advert-story} illustrates our system. Our system automatically detects the presence of a billboard in an image frame from the video sequence. Post billboard detection, our system also localizes its position in the image frame. The user is given an opportunity to manually adjust and refine the detected four corners of the billboard. Finally, a new advertisement is integrated into the image, and tracked across all frames of the video sequence. Thereby, we generate a new composite video with the integrated advert.

Currently, there are no such existing framework available in the literature that aid the marketing agents to seamlessly integrate a new advertisement, into an original video sequence. However, a few companies viz.\ Mirriad~\cite{mirriad2018} uses patented advertisement plantation technique to integrate 3D objects in a video sequence.  

\vspace{-0.2in}

\section{Technology}
\vspace{-0.1in}
The backbone of our advert creation system is based on state-of-the-art techniques from deep learning and image processing. In this section, we briefly describe the underlying techniques used in the various components of the demo system. The different modules of our system are: advert- recognition, localization, and integration. 

\vspace{-0.2in}

\subsection{Advert Recognition}
The first module of our advert creation system is used for the recognition of billboard~\footnote{In this paper, we interchangeably use both the terms, \emph{billboard} and \emph{advert} to indicate a candidate object for new advertisement integration in an image frame.} -- does an image frame from the video sequence contain billboard? This helps the system user to automatically detect the presence of billboard in an image frame of the video. We use a deep neural network (DNN) as a binary classifier where classes represent \emph{presence} and \emph{absence} of billboard in video frame respectively. We use a VGG-based network~\cite{simonyan2014very} for billboard detection. We use transfer learning with pre-trained ImageNet weights. We freeze the corresponding weights of all layers apart from last $5$ layers. We add $3$ fully connected layers with a \emph{softmax} layer as the output layer. We train this deep network on our annotated dataset, containing both billboard and non-billboard images, and achieve good accuracy on billboard recognition.


\vspace{-0.2in}

\subsection{Advert Localization}
\begin{figure}
\vspace{-0.2in}
\centering
\subfloat[Input Image]{\includegraphics[width=0.24\textwidth]{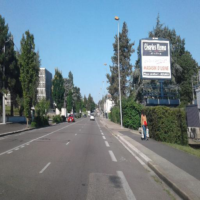}}\ 
\subfloat[Ground Truth]{\includegraphics[width=0.24\textwidth]{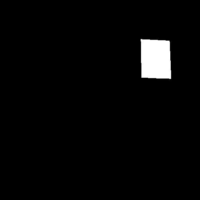}}\ 
\subfloat[Detected Advert]{\includegraphics[width=0.24\textwidth]{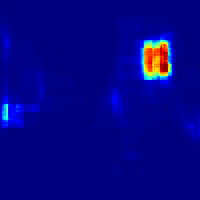}}\ 
\subfloat[Localized Advert]{\includegraphics[width=0.24\textwidth]{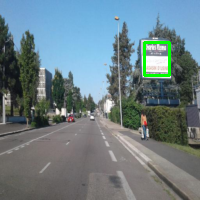}}
\caption{Localization of billboard using our advert creation system. We localize the advert from the probabilistic heatmap, by circumscribing a rectangle with minimum bounding area.}
\label{fig:locate-advert}
\vspace{-0.5in}
\end{figure}

The second module of our advert creation system is used for localizing the position of recognized billboard -- where is the billboard located in image frame? We use a encoder-decoder based deep neural network that localizes the billboard position in an image. We train this model on our billboard dataset comprising input images (cf.\ Fig.\ \ref{fig:locate-advert}(a)) and corresponding binary ground truth image (cf.\ Fig.\ \ref{fig:locate-advert}(b)). We train the model for several thousands of epochs. The localized billboard is a probabilistic image, that denotes the probability of an image pixel to belong to \emph{billboard} class. We generate the binary threshold image from our computed heatmap using thresholding, and detect the various closed contours on the binary image. Finally, we select the contour with the largest area as our localized billboard position. We thereby compute the initial four corners from the binary image by circumscribing a rectangle on the selected contour with minimum bounding area. The localized advert is shown in Fig.\ \ref{fig:locate-advert}.

\subsection{Advert Integration}
The third and final module of our system is advert integration -- how to integrate a new advert in the video? In this stage, the localized billboard is replaced with a new advert in a seamless and temporally consistent manner. We use Poisson image editing~\cite{perez2003poisson} on the new advert, to achieve similar local illumination and local color tone, as the original video sequence. Furthermore, the relative motion of the billboard within the scene is tracked using Kanade-Lucas-Tomasi (KLT)~\cite{lucas1981an} tracking technique.

\vspace{-0.2in}
\section{Design and Interface}
We have designed an online system to demonstrate the functionalities of the various modules~\footnote{A demonstration video of our advert creation system can be accessed via \url{https://youtu.be/zaKpJZhBVL4}}. The web UI interface is designed in \texttt{Vue.js} - the progressive JavaScript Framework. The back end is supported via \texttt{Express} - Node.js web application framework. The deep neural networks for advert recognition and localization is designed in pure \texttt{python}, and the advert integration is implemented in \texttt{C++}. The web service to support advert detection is performed in \texttt{python flask}. The integration of a new advert into the existing video in the web server is executed via \texttt{C++} binary.

\begin{figure}[htb]
\vspace{-0.2in}
\centering
\includegraphics[width=0.88\textwidth]{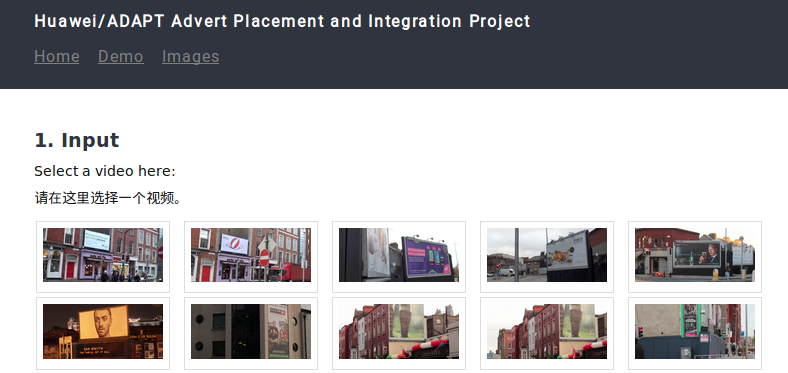}
\caption{Interface of the demo for advert detection and integration.}
\label{fig:tool-interface}
\vspace{-0.3in}
\end{figure}

Figure~\ref{fig:tool-interface} illustrates a sample snapshot of our developed web-based tool. The web interface consists of primarily three sections: \texttt{Home}, \texttt{Demo} and \texttt{Images}. The page \texttt{Home} provides an overview of the system. The next page \texttt{Demo} describes the entire working prototype of our system. The user selects a sample video from the list, runs the billboard detection module to accurately localize the billboard at sample image frames of the video. The detection module estimates the four corners of the billboard. However, the user also gets an option to \emph{refine} the four corners manually, if the detected four corners are not completely accurate. The refined four corners of the billboard are subsequently used for tracking and integration of a new advertisement into the video sequence. The third and final web page \texttt{Images} contains the list of all candidate adverts that can be integrated into the selected video sequence.

Finally, our system integrates the new advertisement into the detected billboard position, and generates a new composite video with the implanted advertisement. 
\vspace{-0.1in}

\section{Conclusion and Future Work}
\vspace{-0.1in}
In this paper, we have presented an online advert creation system on multimedia videos for a personalized and targeted advertisement. We use techniques from deep neural networks and image processing, for a seamless integration of new adverts into existing videos. Our system is trained on datasets that comprises outdoor scenes and views. Our future work involve further refining the performance of the system, and also generalizing it to other video sequence types.

\small{
\section*{Acknowledgement}
\vspace{-0.1in}
The ADAPT Centre for Digital Content Technology is funded under the SFI Research Centres Programme (Grant 13/RC/2106) and is co-funded under the European Regional Development Fund.
}

\vspace{-0.2in}
%
%

\end{document}